\documentclass[12pt]{article}
\def\nabstar#1{\nabla\kern-0.5pt\smash{\raise 4.5pt\hbox{$\ast$}}
               \kern-4.5pt_{#1}}

\def\drvstar#1{\partial\kern-0.5pt\smash{\raise 4.5pt\hbox{$\ast$}}
               \kern-5.0pt_{#1}}

\def\newline{\relax\ifhmode\null\hfil\break\else\nonhmodeerr@\newline\fi}
\def\frac#1#2{{#1\over#2}}
\def\text#1{{\hbox{\rm #1}}}
\def\flushpar{{\par \noindent}}

\newcommand{\beq}{\begin{equation}}
\newcommand{\eeq}{\end{equation}}
\newcommand{\bea}{\begin{eqnarray}}
\newcommand{\eea}{\end{eqnarray}}
\def\Id{ \mbox{1\hspace{-1.2mm}I} }
\def\BE{\begin{equation}}
\def\EE{\end{equation}}
\def\BA{\begin{eqnarray}}
\def\EA{\end{eqnarray}}
\def\BAN{\begin{eqnarray*}}
\def\EAN{\end{eqnarray*}}

\def\nn{\nonumber\\}

\def\tr{\mbox{tr}}

\def\det{\mbox{det}}

\def\gm5{\gamma_5}

\input epsf.sty
\newdimen\psfigsize
\def\psfigure#1 #2 #3 #4 #5{
    \begin{figure}[tbh]
      \begin{center}
      \vbox{
        \null\vskip-0.2in\hskip#2
        \epsfxsize=#1
        \epsfbox{#4}
        \vskip -0.3in
        \caption {#5 \label{#3}}
        \vskip 0.0 true in plus 0.3 true in
      }
      \end{center}
   \end{figure}
}
\usepackage{graphicx}
\begin{document}
\thispagestyle{empty}
\begin{flushright}
NTUTH-03-505C \\
May 2003
\end{flushright}
\bigskip\bigskip\bigskip
\vskip 0.5truecm
\begin{center}
{\LARGE {
Light quark masses, chiral condensate and quark-gluon condensate
in quenched lattice QCD with exact chiral symmetry}}
\end{center}
\vskip 0.5truecm
\centerline{Ting-Wai Chiu and Tung-Han Hsieh}
\vskip5mm
\centerline{Department of Physics, National Taiwan University}
\centerline{Taipei, Taiwan 106, Taiwan.}
\centerline{\it E-mail : twchiu@phys.ntu.edu.tw}
\vskip 0.5truecm
\bigskip \nopagebreak \begin{abstract}
\noindent

We determine several quantities in quenched lattice QCD with
exact chiral symmetry. For 100 gauge configurations generated with
Wilson gauge action at $ \beta = 6.0 $ on the $ 16^3 \times 32 $ lattice,
we compute quenched quark propagators for 13 bare quark masses.
The pion decay constant is extracted from the pion propagator, and
from which the inverse lattice spacing is determined to be
$ a^{-1} = 1.979(6) $ GeV.
The parameters ($ \delta, A, B $) in the pseudoscalar meson mass
formula (\ref{eq:mpi2_c}) in quenched chiral perturbation theory (q$\chi$PT)
to one-loop order are determined.
Further, we measure the index (topological) susceptibility
of these 100 gauge configurations, $ \chi_t = (175 \pm 6 \mbox{ MeV} )^4 $,
from which we obtain an estimate of the mass of $ \eta' $ in q$\chi$PT,
and the coefficient of quenched chiral logarithm,
both in good agreement with the values determined from
the pion masses, as well as with the theoretical estimates.
With our values of $ \delta, A, B $, the experimental inputs of pion and kaon
masses, and the pion decay constant, we determine the light quark masses:
$ m_{u,d} = 4.1 \pm 0.3 $ MeV, and $ m_s = 92 \pm 9 $ MeV,
in the $ \overline{\mbox{MS}} $ scheme at scale $ \mu = 2 $ GeV.
Also, we determine the quark condensate $ \langle \bar q q \rangle
= -(250 \pm 3 \mbox{ MeV})^3 $,
and the quark-gluon condensate
$ g \langle \bar q \sigma_{\mu\nu} F_{\mu\nu} q \rangle
= -(434 \pm 4 \mbox{ MeV})^5 $,
in the $ \overline{\mbox{MS}} $ scheme at scale 2 GeV.

\vskip 0.5cm
\noindent PACS numbers: 11.15.Ha, 11.30.Rd, 12.38.Gc, 11.30.Fs

\noindent Keywords : Quark Mass, Chiral Condensate, Quark-Gluon Condensate,
Lattice QCD, Domain-Wall quark, Chiral Perturbation Theory

\end{abstract}
\vskip 1.5cm

\newpage\setcounter{page}1

\section{Introduction}

Exact chiral symmetry on the lattice was pioneered by
Kaplan \cite{Kaplan:1992bt} with his proposal of domain-wall fermions
(DWF) on the 5-dimensional lattice, in which the fifth dimension
(internal flavor space)
is discretized with $ N_s $ sites (flavors) and lattice spacing $ a_5 $.
Although the initial motivation was to provide a nonperturbative formulation
of chiral gauge theories, the idea turns out to be
natural and well-defined for vector-like gauge theories (e.g. QCD), with
quark fields constructed from the boundary modes with open boundary
conditions \cite{Shamir:1993zy}.
Soon after Kaplan proposed DWF for chiral gauge theories,
Narayanan and Neuberger \cite{Narayanan:wx} observed that the
chiral determinant can be
written as the inner-product (``overlap'') of two fermionic many body
states of two bilinear Hamiltonians.
For vector gauge theories like QCD, the fermion determinant
is the product of a complex conjugate pair of chiral determinants,
thus it is gauge invariant, real, non-negative, and the corresponding
lattice Dirac operator for massless quarks
(i.e., the overlap Dirac operator \cite{Neuberger:1998fp})
can be represented by a finite matrix of fixed shape regardless of the
topology of the background gauge field, without undesired doubling or
any fine-tuning.
Mathematically, the overlap Dirac operator is
exactly equal to the effective 4D lattice Dirac operator (for internal
fermions dressed with pseudofermions) of DWF (with $m_q=0$)
in the limit $ N_s \to \infty $ followed by $ a_5 \to 0 $,
\bea
\label{eq:D}
D = m_0 \left( 1 + \gamma_5 \frac{H_w}{\sqrt{H_w^2}} \right) \ ,
\hspace{4mm} H_w = \gamma_5 D_w
\eea
where $ D_w $ is the standard Wilson Dirac operator plus a
negative parameter $ -m_0 $ ($ 0 < m_0 < 2 $).


However, it has been shown \cite{Chiu:2002ir} that the conventional DWF 
with any finite $ N_s $ does {\it not} preserve the chiral symmetry optimally 
for any gauge background. 

Recently, one of us (T.W.C.) has proposed a new DWF action
\cite{Chiu:2002ir,Chiu:2003ir}
such that the quark propagator preserves
the chiral symmetry optimally for any $ N_s $ and background
gauge field. Further, its effective 4D lattice Dirac operator
(with any $ N_s $ and $ m_q $) for internal fermion loops
is exponentially-local for sufficiently smooth gauge backgrounds
\cite{Chiu:2002kj}. In the optimal DWF, the quark fields are
constructed from the boundary modes at $ s=0 $ and $ s=N_s + 1 $
\cite{Chiu:2003ir},
\bea
\label{eq:q}
q(x) &=& \sqrt{r} \left[ P_{-} \psi(x,0) + P_{+} \psi(x,N_s+1) \right] \ ,
\hspace{4mm} r = \frac{1}{2 m_0} \\
\label{eq:qbar}
\bar q(x) &=& \sqrt{r}
\left[ \bar\psi(x,0) P_{+} + \bar\psi(x,N_s+1) P_{-} \right] \ ,
\eea
and the generating functional for $n$-point Green's function
of the quark fields has been derived \cite{Chiu:2003ir},
\bea
\label{eq:ZW_odwf}
Z[J,\bar J] =
\frac{\int [dU] e^{-{\cal A}_g} \det [(D_c+m_q)(1+rD_c)^{-1}]
           \exp \left\{ \bar J (D_c+ m_q)^{-1} J \right\}  }
     {\int [dU] e^{-{\cal A}_g} \det [(D_c + m_q)(1+r D_c)^{-1}] }
\eea
where $ {\cal A}_g $ is the action of the gauge fields,
$ m_q $ is the bare quark mass, $ \bar J $ and $ J $
are the Grassman sources of $ q $ and $ \bar q $ respectively, and
\bea
\label{eq:Dc}
r D_c &=& \frac{ 1 + \gamma_5 S_{opt} }{ 1 - \gamma_5 S_{opt} } \ , \\
S_{opt} &=&
  \left\{ \begin{array}{ll}
           H_w R_Z^{(n,n)}(H_w^2),   &  N_s = 2n+1,   \\
           H_w R_Z^{(n-1,n)}(H_w^2), &  N_s = 2n.    \\
          \end{array} \right.
\label{eq:S_opt_RZ}
\eea
Here $ R_Z(H_w^2) $ is the Zolotarev optimal rational polynomial
\cite{Akhiezer:1992} for the inverse square root of $ H_w^2 $,
\bea
\label{eq:rz_nn}
R^{(n,n)}_Z(H_w^2) &=& \frac{d_0}{\lambda_{min}}
\prod_{l=1}^{n} \frac{ 1+ h_w^2/c_{2l} }{ 1+ h_w^2/c_{2l-1} } \nn
&=& \frac{1}{\lambda_{min}} (h_w^2 + c_{2n})
    \sum_{l=1}^n \frac{b_l}{h_w^2 + c_{2l-1}} \ ,
\hspace{8mm}  h_w^2 = H_w^2/\lambda_{min}^2
\eea
and
\bea
\label{eq:rz_n1n}
R^{(n-1,n)}_Z(H_w^2) = \frac{d'_0}{\lambda_{min}}
\frac{ \prod_{l=1}^{n-1} ( 1+ h_w^2/c'_{2l} ) }
     { \prod_{l=1}^{n} ( 1+ h_w^2/c'_{2l-1} ) }
= \frac{1}{\lambda_{min}} \sum_{l=1}^n \frac{b'_l}{h_w^2 + c'_{2l-1}} \ ,
\eea
where the coefficients $ d_0 $, $ d'_0 $, $ c_l $ and $ c'_l $
are expressed in terms of elliptic functions \cite{Akhiezer:1992}
with arguments depending only on $ N_s $
and $ \lambda_{max}^2 / \lambda_{min}^2 $
($ \lambda_{max} $ and $ \lambda_{min} $ are the maximum and the minimum
of the the eigenvalues of $ | H_w | $).

From (\ref{eq:ZW_odwf}), the effective 4D lattice Dirac operator
for the fermion determinant is
\bea
\label{eq:Dm_odwf}
D(m_q) = (D_c+m_q)(1+rD_c)^{-1}
       = m_q + (m_0 - m_q/2) \left[ 1 + \gamma_5 H_w R_Z(H_w^2) \right] \ ,
\eea
and the quark propagator in background gauge field is
\bea
\label{eq:quark_prop}
\langle q(x) \bar q(y) \rangle &=&
- \left. \frac{\delta^2 Z[J,\bar J]}{\delta \bar J(x) \delta J(y)}
\right|_{J=\bar J=0}                                             \nn
&=& (D_c + m_q)^{-1}_{x,y}=(1-rm_q)^{-1}[D^{-1}_{x,y}(m_q) - r \delta_{x,y}]
\eea

In practice, we have two ways to evaluate the
quark propagator (\ref{eq:quark_prop}) in background gauge field: \\
(i) To solve the linear system of the 5D optimal
DWF operator, as outlined in Ref. \cite{Chiu:2003ir}; \\
(ii) To solve $ D^{-1}_{x,y}(m_q) $ from the system
\bea
\label{eq:Dm_RZ}
D(m_q) Y =
\left[m_q+(m_0-m_q/2) \left(1+\gamma_5 H_w R_Z(H_w^2)\right) \right] Y
= \Id \ ,
\eea
with nested conjugate gradient \cite{Neuberger:1998my}, and then
substitute the solution vector $ Y $ into (\ref{eq:quark_prop}).

Since either (i) or (ii) yields exactly the same quark propagator,
in principle, it does {\it not} matter which linear system one actually
solves. However, in practice, one should pick the most efficient scheme
for one's computational system (hardware and software).

For our present software and hardware
(a Linux PC cluster of 42 nodes \cite{Chiu:2002bi}), it has been
geared to the scheme (ii), and it attains the maximum efficiency
with Neuberger's 2-pass algorithm \cite{Neuberger:1998jk}
for the inner conjugate gradient loop of (\ref{eq:Dm_RZ}).
So, in this paper, we use the scheme (ii) to compute the quark propagator,
with the quark fields (\ref{eq:q})-(\ref{eq:qbar})
defined by the boundary modes in the optimal DWF.
A study of Neuberger's 2-pass algorithm is   
presented in Ref. \cite{Chiu:2002fy}.

For our computations in this paper, we fix $ m_0 = 1.3 $, and project
out 20 low-lying eigenmodes and two high-lying modes of $ |H_w| $.
With $ N_s = 32 $, i.e., 16-th order Zolotarev rational 
polynomial $ R_Z^{(15,16)}(H_w^2) $ in (\ref{eq:Dm_RZ}), we impose 
stopping criteria $ 2 \times 10^{-12} $ and $ 10^{-11} $
for the inner and outer conjugate gradient loops respectively.
Then the chiral symmetry breaking due to finite $ N_s $ is always
less than $ 10^{-13} $,
\bea
\sigma = \left| \frac{Y^{\dagger} S_{opt}^2 Y}{Y^{\dagger} Y} - 1 \right|
< 10^{-13} \ ,
\eea
for every iteration of the nested conjugate gradient.

The outline of this paper is as follows.
In Section 2, we determine the parameters ($ \delta, A, B $) in the
pseudoscalar meson mass formula (\ref{eq:mpi2_c}) in quenched chiral
perturbation theory (q$\chi$PT) to one-loop order,
by computing quenched quark propagators for 13 bare
quark masses, and for 100 gauge configurations generated with
Wilson gauge action at $ \beta = 6.0 $ on the $ 16^3 \times 32 $ lattice.
The pion decay constant is extracted from the pion propagator, and
from which the lattice spacing is determined to be
$ a $ = 0.0997(3) fm = 0.5053(15) $ \mbox{GeV}^{-1} $.
Also, we measure the index (topological) susceptibility of these
100 gauge configurations, $ \chi_t = (175 \pm 6 \mbox{ MeV} )^4 $,
which provides an estimate of
$ m_{\eta'} = 813 \pm 51 \mbox{ MeV} $ in q$\chi$PT,
and the coefficient of quenched chiral logarithm $ \delta = 0.16(2) $,
which is consistent with $\delta = 0.164(13) $ determined from
the pion masses.
In Section 3, with our determined values of $ \delta, A, B $,
the experimental inputs of pion and kaon masses,
and the pion decay constant, we determine the light quark masses:
$ m_{u,d} = 4.1 \pm 0.3 $ MeV, and $ m_s = 92 \pm 9 $ MeV,
in the $ \overline{\mbox{MS}} $ scheme at scale $ \mu = 2 $ GeV.
In Section 4, we determine the quark condensate
$ \langle \bar q  q \rangle = -(250 \pm 3 \mbox{ MeV})^3 $,
and the quark-gluon condensate
$ g \langle \bar q \sigma_{\mu\nu} F_{\mu\nu} q \rangle
= -(434 \pm 4 \mbox{ MeV})^5 $,
in the $ \overline{\mbox{MS}} $ scheme at scale 2 GeV.

\section{Determination of the Parameters in
Quenched Chiral Perturbation Theory}

In quenched chiral perturbation theory (q$\chi$PT)
\cite{Sharpe:1992ft,Bernard:1992mk}
the pion mass to one-loop order reads
\bea
\label{eq:mpi2}
m_\pi^2 = C m_q \{ 1 - \delta[ \mbox{ln}( C m_q/\Lambda_{\chi}^2 ) + 1 ] \}
          + B m_q^2  
\eea
where $ m_q $ denotes the bare ($ u $ and $ d $) quark mass,
$ \Lambda_{\chi} $ is the chiral cutoff,
$ C $ and $ B $ are parameters,
and $ \delta $ is the coefficient of the quenched chiral logarithm. 
Further, summing all diagrams in which each internal $ \eta' $ propagator
is decorated with any number of ``bubble sums''(i.e., summing the leading 
$ [\delta \ln(C m_q/\Lambda_\chi^2)]^n $ terms to all orders), 
one obtains \cite{Sharpe:1992ft}  
\bea
\label{eq:mpi2_c}
m_\pi^2 = A m_q^{\frac{1}{1+\delta}} + B m_q^2
\eea
where 
\bea
\label{eq:A}
A = \left[ C \left(\frac{\Lambda_\chi^2}{e} \right)^{\delta} 
    \right]^{\frac{1}{1+\delta}} \ . 
\eea
Obviously, the value of $ \delta $ is independent of the chiral 
cutoff $ \Lambda_\chi $ (which is usually taken to be 
$ 2 \sqrt{2} \pi f_{\pi} $ with $ f_{\pi} \simeq 132 $ MeV).

In this section, we determine $ \delta $ by fitting our data of 
$ m_\pi^2 $ to (\ref{eq:mpi2}) and (\ref{eq:mpi2_c}) respectively, 
and find that these two values of $ \delta $ are in good agreement 
with each other.

Theoretically, $ \delta $ can be estimated as \cite{Sharpe:1992ft}
\bea
\label{eq:delta}
\delta = \frac{m_{\eta'}^2}{8 \pi^2 f_\pi^2 N_f} \ ,
\eea
where $ f_\pi $ denotes the pion decay constant,
$ N_f $ denotes the number of light quark flavors, and
$ m_{\eta'} $ denotes the $ \eta' $ mass in q$\chi$PT,
which can be estimated as
\bea
\label{eq:eta_cpt}
\sqrt{ {\bf m}_{\eta'}^2 + m_{\eta}^2 - 2 m_{K}^2 } = 853  \mbox{ MeV } \ ,
\eea
with experimental values of meson masses\footnote{Here
we distinguish the physical $ \eta' $ mass ($ {\bf m}_{\eta'} $),
from the $ \eta' $ mass ($ m_{\eta'} $) in q$\chi$PT.}
: $ {\bf m_{\eta'} } = 958 $ MeV, $ m_{\eta} = 547 $ MeV,
and $ m_{K} = 495 $ MeV.
For $f_{\pi}=132$ MeV, $N_f=3 $, and $ m_{\eta'}= 853 $ MeV,
(\ref{eq:delta}) gives
\bea
\label{eq:delta_cpt}
\delta \simeq 0.176 \ .
\eea

Evidently, if one can extract $ \delta $ from the data of $ m_\pi^2 $,
then $ m_{\eta'} $ in q$\chi$PT can be determined by (\ref{eq:delta}).

Besides from the data of $ m_{\pi}^2 $,
one can also obtain $ \delta $ via (\ref{eq:delta})
by extracting $ m_{\eta'} $ from the propagator
of the disconnected hairpin diagram.
However, to compute the propagator of the hairpin is very tedious.

Fortunately, with the realization of exact chiral symmetry on the
lattice, the quark propagator coupling to
$ \eta' $ is $ ( D_c + m_q )^{-1} $,
thus only the zero modes of $ D = D_c ( 1 + r D_c )^{-1} $ can contribute
to the hairpin diagram, regardless of the bare quark mass $ m_q $.
Therefore one can derive an exact relation
\cite{Chiu:2002xm,DeGrand:2002gm,Giusti:2001xh}
between the $ \eta' $ mass in q$\chi$PT and the index susceptibility of $D$,
without computing the hairpin diagram at all.
Explicitly, this exact relation reads as
\bea
\label{eq:exact_wv}
( m_{\eta'} a )^2
= \frac{4 N_f }{(f_{\pi}a)^2}
  \frac{\langle (n_{+}-n_{-})^2 \rangle }{N}
= \frac{4 N_f }{(f_{\pi}a)^2} \chi_t \ ,
\eea
where $ N $ is the total number of sites, and
$ \chi_t \equiv \langle (n_{+} - n_{-})^2 \rangle / N $ is the
index susceptibility of $ D $ in the quenched approximation.
Then (\ref{eq:delta}) and (\ref{eq:exact_wv}) together gives
\bea
\label{eq:delta_s}
\delta
= \frac{ 1 }{ 2 {\pi}^2 ( f_{\pi} a )^4 }
  \frac{ \langle ( n_{+} - n_{-} )^2 \rangle }{N}
\eea
A salient feature of (\ref{eq:delta_s}) is that $ \delta $
can be determined at finite lattice spacing $ a $, by measuring
the index (susceptibility) of $ D $, and
with $ f_{\pi} a $ extracted from the pion propagator.

Now it is clear that, in order to confirm the presence of
quenched chiral logarithm in lattice QCD,
one needs to check whether the coefficient $ \delta $
obtained by fitting (\ref{eq:mpi2_c}) to the data of $ m_{\pi}^2 $,
agrees with that (\ref{eq:delta_s}) from the index susceptibility.
This is a requirement for the consistency of
the theory, since the quenched chiral logarithm in $ m_{\pi}^2 $
(\ref{eq:mpi2}) is due to the $ \eta' $ loop coupling to the
pion propagator through the mass term (in the chiral lagrangian),
thus $ \delta $ ($ m_{\eta'} $) must be the same in both cases.
We regard this consistency requirement as a basic criterion for
lattice QCD (with any fermion scheme) to realize QCD chiral
dynamics in continuum.

Recently, we have determined $\delta=0.203(14)$, from the data of
$ m_\pi^2 $, as well as $\delta=0.197(27)$ from the index susceptibility,
for the $ 8^3 \times 24 $ lattice at $ \beta = 5.8 $ \cite{Chiu:2002xm}.
Their good agreement suggests that lattice QCD with exact chiral symmetry
indeed realizes QCD chiral dynamics in the continuum.

In this paper, we extend our studies to a lattice of larger
volume and smaller lattice spacing, $ 16^3 \times 32 $ at
$ \beta = 6.0 $. Details of our computational
scheme have been given in Ref. \cite{Chiu:2002xm}, except
the inner conjugate gradient loop is now iterated with Neuberger's
2-pass algorithm \cite{Neuberger:1998jk} which not only provides very
high precision of chiral symmetry with a fixed amount of memory,
but also increases ($\sim20\% $) the speed of our
computational system (a Linux PC cluster
of 42 nodes at the speed $\sim50 $ Gflops \cite{Chiu:2002bi}).
A study of Neuberger's 2-pass algorithm
is presented in Ref. \cite{Chiu:2003ub}.

We retrieved 100 gauge configurations from the repository at
the Gauge Connection (http://qcd.nersc.gov),
which were generated with Wilson gauge action
at $ \beta = 6.0 $ on the $ 16^3 \times 32 $ lattice \cite{Kilcup:1996hp}.
For each configuration, quark propagators are computed
for 13 bare quark masses.

Then the pion propagator
\bea
\label{eq:pion}
M_{5}(\vec{x},t;\vec{0},0) =
\tr \{ \gamma_5 ( D_c + m_q )^{-1} ( 0, x ) \gamma_5
                ( D_c + m_q )^{-1} ( x, 0 ) \} \ ,
\eea
(where the trace runs over the Dirac and color space),
and its time correlation function
\bea
\label{eq:Gt}
G(t) = \sum_{ \vec{x} } M_{5}(\vec{x},t;\vec{0},0) \ ,
\eea
are obtained, and $ \langle G(t) \rangle $ is fitted by the usual
formula
\BAN
\label{eq:Gt_fit}
G_{\pi}(t) = \frac{Z}{2 m_{\pi} a }
             [ e^{-m_{\pi} a t} + e^{-m_{\pi} a (T-t)} ]
\EAN
to extract the pion mass $ m_{\pi} a $ and the pion decay constant
\BAN
\label{eq:fpi}
f_{\pi} a = 2 m_q a \frac{\sqrt{Z}}{m_{\pi}^2 a^2 } \ .
\EAN

\begin{figure}[htb]
\begin{center}
\hspace{0.0cm}\includegraphics*[height=12cm,width=10cm]{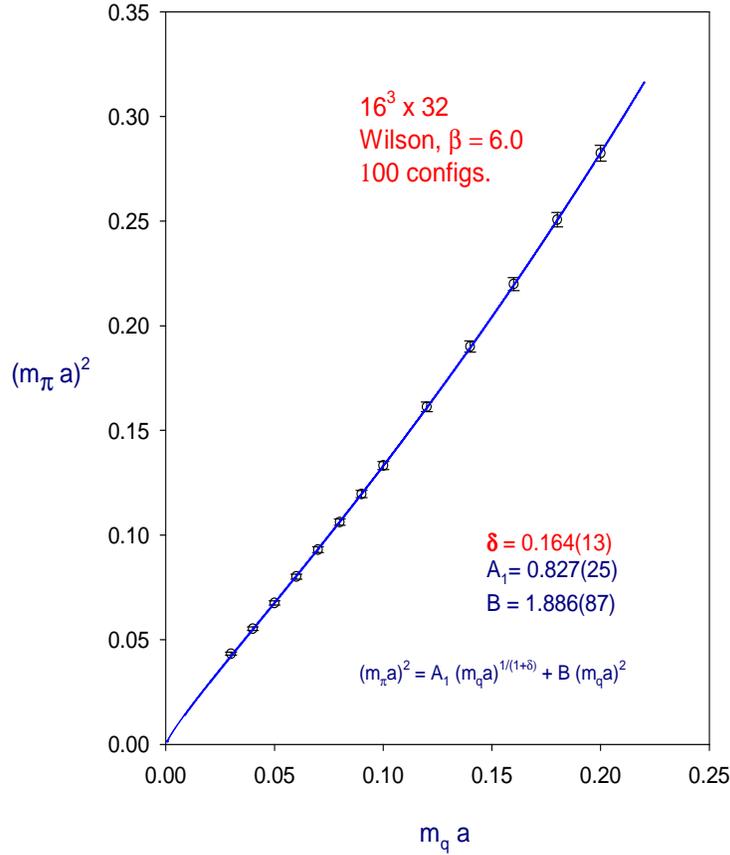}
\caption{The pion mass square $ (m_\pi a)^2 $ versus the
bare quark mass $ m_q a $. The solid line is the fit of Eq. (\ref{eq:mpi2_c}).}
\label{fig:mpi2_16c}
\end{center}
\end{figure}

\begin{figure}[htb]
\begin{center}
\hspace{0.0cm}\includegraphics*[height=10cm,width=10cm]{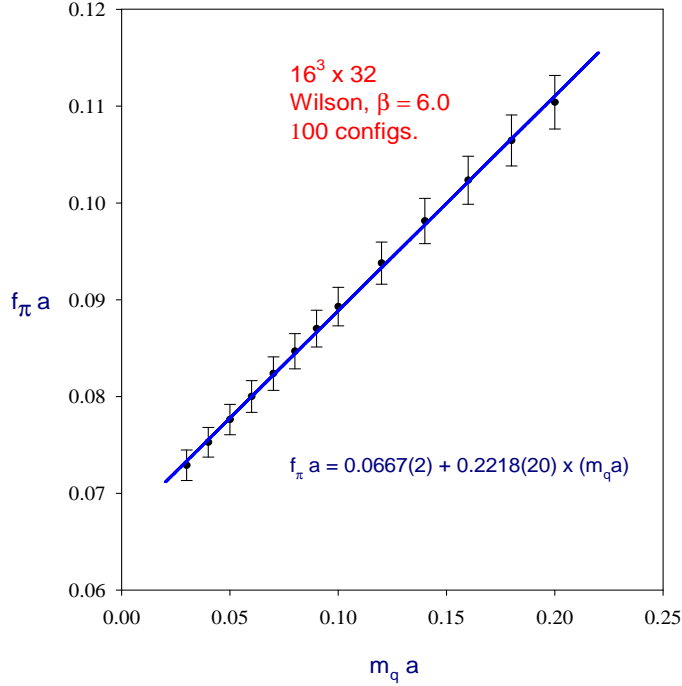}
\caption{The pion decay constant $ f_\pi a $ versus the
bare quark mass $ m_q a $. The solid line is the linear fit.}
\label{fig:fpi_16}
\end{center}
\end{figure}

\begin{figure}[htb]
\begin{center}
\hspace{0.0cm}\includegraphics*[height=12cm,width=10cm]{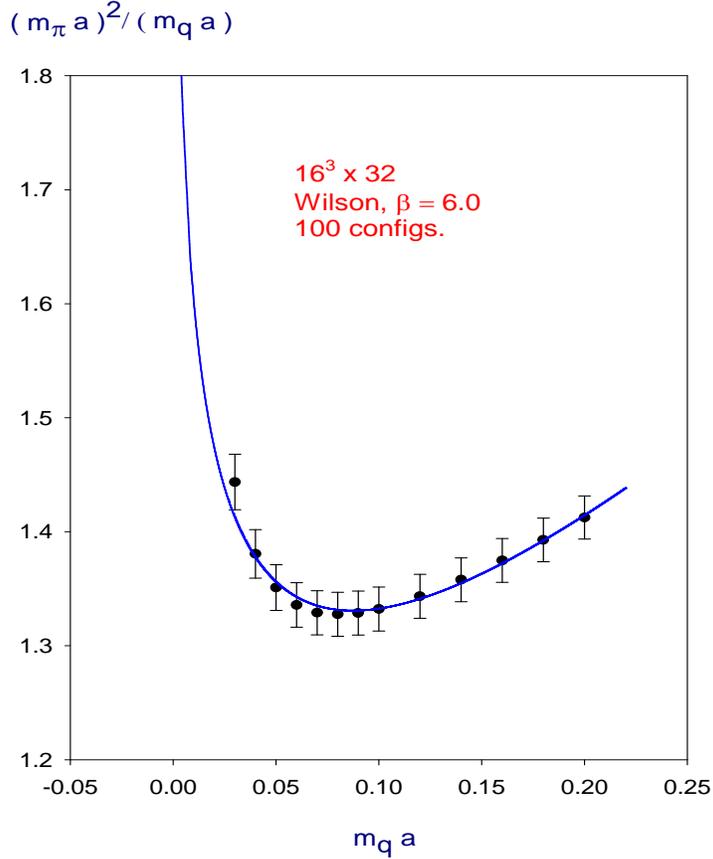}
\caption{$(m_\pi a)^2 / (m_q a) $ versus the bare quark mass $ m_q a $.}
\label{fig:mpi2omq_c}
\end{center}
\end{figure}

\begin{figure}[htb]
\begin{center}
\hspace{0.0cm}\includegraphics*[height=12cm,width=10cm]{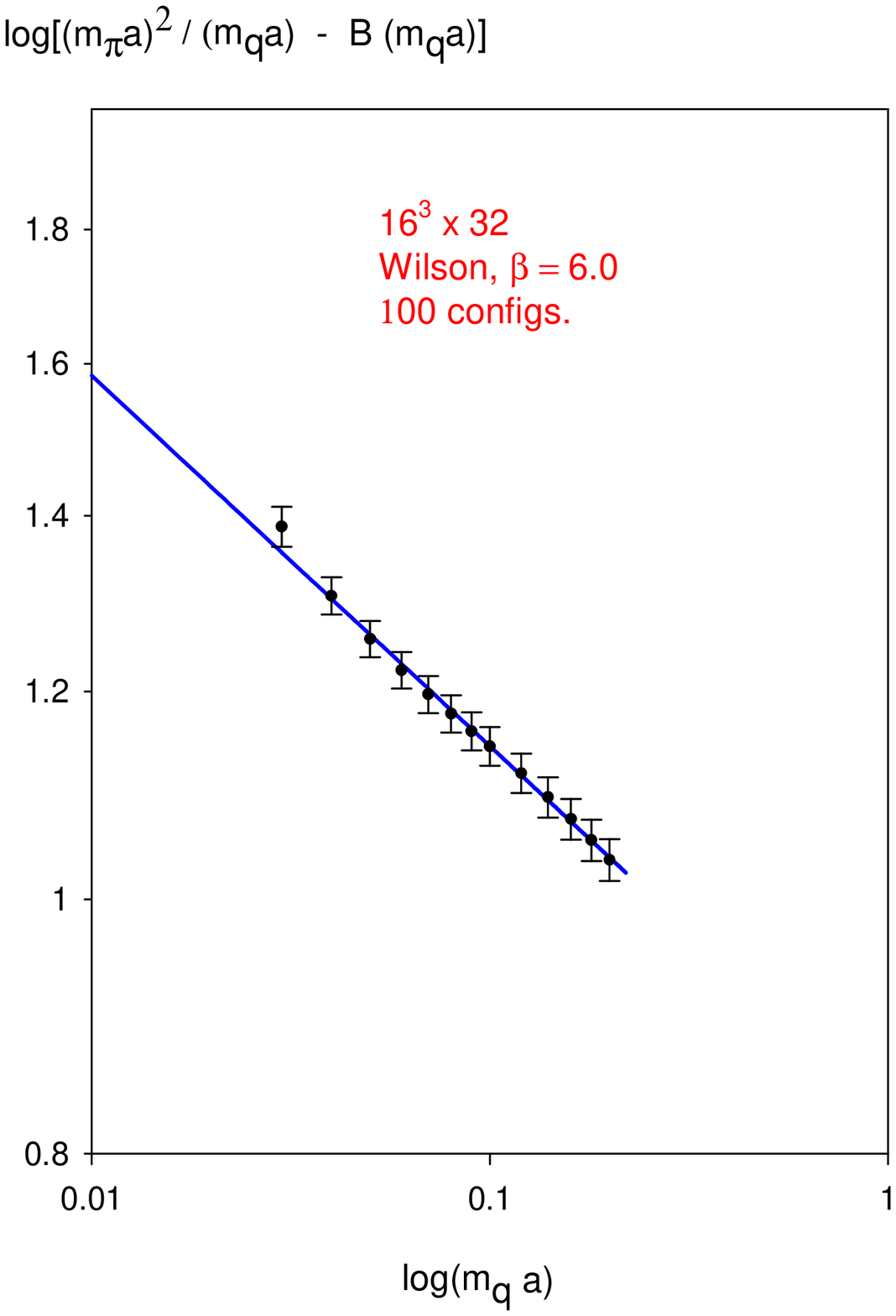}
\caption{The extraction of the quenched chiral logarithm by plotting
$ \log[(m_\pi a)^2 / (m_q a) - B (m_q a)] $ versus $ \log(m_q a) $. 
The slope of the fitted straight line is equal to $ -\delta/(1+\delta) $.}
\label{fig:chiralog_c}
\end{center}
\end{figure}

In Figs. \ref{fig:mpi2_16c} and \ref{fig:fpi_16},
we plot the pion mass square $ (m_\pi a)^2 $ and decay constant
$ f_\pi a $ versus bare quark mass $ m_q a $, respectively.

The data of $ f_{\pi} a $ (see Fig. \ref{fig:fpi_16}) can be fitted by
\BAN
f_{\pi} a = 0.0667(2) + 0.2218(20)  \times  ( m_q a )
\EAN
with $\chi^2$/d.o.f.=0.95. Thus taking $ f_\pi a $ at $ m_q a = 0 $
equal to $ 0.132 $ GeV times the lattice spacing $ a $,
we can determine the lattice spacing $ a $ and its inverse,
\bea
a^{-1} &=& \frac{0.132}{f_0} \mbox{ GeV} = 1.979(6) \mbox{ GeV} \ , \nn
a &=& 0.0997(3) \mbox{ fm} \ .
\label{eq:a}
\eea
Thus the size of the lattice is
$ \sim(1.6 \mbox{ fm})^3 \times 3.2 \mbox{ fm} \simeq (1.9 \mbox{ fm})^4 $.
Since the smallest pion mass is $ 412 \mbox{ MeV} $, the lattice
size is $ \sim(3.3)^3 \times 6.6 $, in units of the Compton wavelength
($\sim0.48 \mbox{ fm}$) of the smallest pion mass. For the largest pion
mass (1052 MeV), the lattice spacing is about 0.5 of its Compton wavelength.

Next, rewriting (\ref{eq:mpi2_c}) as 
\BAN
\label{eq:mpi2_c1}
m_\pi^2 a^2 = A_1 (m_q a)^{\frac{1}{1+\delta}} + B (m_q a)^2, \hspace{4mm}  
A_1 = A a^{-\frac{1-2\delta}{1+\delta}}
\EAN
and fitting to the data of $ m_\pi^2 $ (see Fig. \ref{fig:mpi2_16c}), 
we obtain
\bea
\label{eq:delta_pi_c} 
\delta &=& 0.164(13) \\
\label{eq:A_c}
A_1 &=& 0.827(25) \\
\label{eq:B_c}
B &=& 1.886(87) 
\eea
with $\chi^2 $/d.o.f.=0.54.

Even though the quenched chiral logarithm may not be easily
detected in the graph of $ (m_\pi a)^2 $ versus $ m_q a $,
it can be unveiled by plotting $ (m_\pi a)^2/(m_q a) $ versus
$ m_q a $, as shown in Fig. \ref{fig:mpi2omq_c}.
Further, plotting $ \log[(m_\pi a)^2/(m_q a)-B(m_q a)] $ versus
$ \log(m_q a) $, the presence of quenched chiral logarithm is evident,
as shown in Fig. \ref{fig:chiralog_c}.

Evidently, the coefficient of quenched chiral logarithm
$ \delta = 0.164(13) $ is in good agreement with the theoretical
estimate $ \delta = 0.176 $ [Eq. (\ref{eq:delta_cpt})] in q$\chi$PT.
Also, from (\ref{eq:delta}), we obtain an estimate of the $ \eta' $ mass
in q$\chi$PT,
\bea
m_{\eta'} = 823 \pm 33 \mbox{ MeV} \ ,
\eea
which is in good agreement with the theoretical estimate (\ref{eq:eta_cpt}).

At this point, it may be instructive to find out what is the value of
$ \delta $ if one fits (\ref{eq:mpi2}) to the data of $ (m_\pi a)^2 $, 
with a fixed $ \Lambda_\chi $, say   
$ \Lambda_{\chi} a = 2 \sqrt{2} \pi f_\pi a =
2 \sqrt{2} \pi \times 0.0667 $. 
The results (with $\chi^2 $/d.o.f.=0.87) are
\bea
\label{eq:delta_pi}
\delta &=& 0.168(17),    \\
\label{eq:Ca}
C a &=& 1.093(20),      \\
\label{eq:B}
B &=& 2.092(125),
\eea
where the value of $ \delta $ (\ref{eq:delta_pi}) is in good agreement 
with (\ref{eq:delta_pi_c}), and other parameters are also consistent
with those determined with (\ref{eq:mpi2_c}).  

Note that, in Fig. \ref{fig:mpi2omq_c},
the minimum of $ (m_\pi a)^2/(m_q a) $ occurs at
$ m_q a \simeq 0.09 $, which corresponds to
$ m_q \simeq 180 $ MeV for $ a \simeq 0.0997 $ fm.
This is consistent with the data of $ 10^3 \times 24 $,
and $ 12^3 \times 24 $ lattices at $ \beta = 5.8 $ \cite{Chiu:2002fy},
where the minima of $ (m_\pi a)^2/(m_q a) $ occur at
$ m_q \simeq $ 190 and 217 MeV respectively.
That is, the minima of $ (m_\pi a)^2/(m_q a) $ for these three lattices
occur in the range of $ m_q \simeq 200 \pm 20 $ MeV.
This suggests that one should be able to observe the
quenched chiral logarithm behavior (in the plot of $ (m_\pi a)^2/(m_q a) $
versus $ m_q a $) for $ m_q \le 200 $ MeV (i.e., $ m_\pi \le 600 $ MeV).

However, in a recent study of chiral logs in quenched lattice QCD
with Iwasaki gauge action \cite{Dong:2003im}, the location of the
minimum of $ (m_\pi a)^2/(m_q a) $ occurs at $ m_q a \simeq 0.13 $
with $ a \simeq 0.2 $ fm, which corresponds to
$ m_q \simeq 130 $ MeV (i.e., $ m_\pi \simeq 480 $ MeV).
This seems to suggest that the location of the minimum
of $ (m_\pi a)^2/(m_q a) $ depends on the gauge action as
well as the lattice spacing $ a $.  
For a recent review of quenched chiral logarithms of various 
quark actions and gauge actions, see Refs. \cite{Giusti:2002rx,Wittig:2002ux}
and the references therein. 

{\footnotesize
\begin{table}
\begin{center}
\begin{tabular}{|c|c|}
\hline
$  n_{+} - n_{-} $   &   number of configurations    \\
\hline
\hline
     5    &       6   \\
\hline
     4    &       5   \\
\hline
     3    &       6   \\
\hline
     2    &       8  \\
\hline
     1    &      14   \\
\hline
     0    &      10   \\
\hline
    -1    &      15   \\
\hline
    -2    &      15   \\
\hline
    -3    &      10   \\
\hline
    -4    &       4   \\
\hline
    -5    &       3   \\
\hline
    -6    &       3   \\
\hline
    -7    &       1   \\
\hline
\hline
\end{tabular}
\end{center}
\caption{
The distribution of the indices of $ D $
for 100 gauge configurations at $ \beta = 6.0 $ on the
$ 16^3 \times 32 $ lattice.
}
\label{table:index}
\end{table}
}

In order to provide a self-consistency check of the coefficient of quenched
chiral logarithm ($ \delta $) determined from the data of $ m_\pi^2 $,
we measure the index susceptibility
of $ D $ by the spectral flow method,
and then determine $ \delta $ via (\ref{eq:delta_s}).

\begin{figure}[htb]
\begin{center}
\hspace{0.0cm}\includegraphics*[height=10cm,width=10cm]{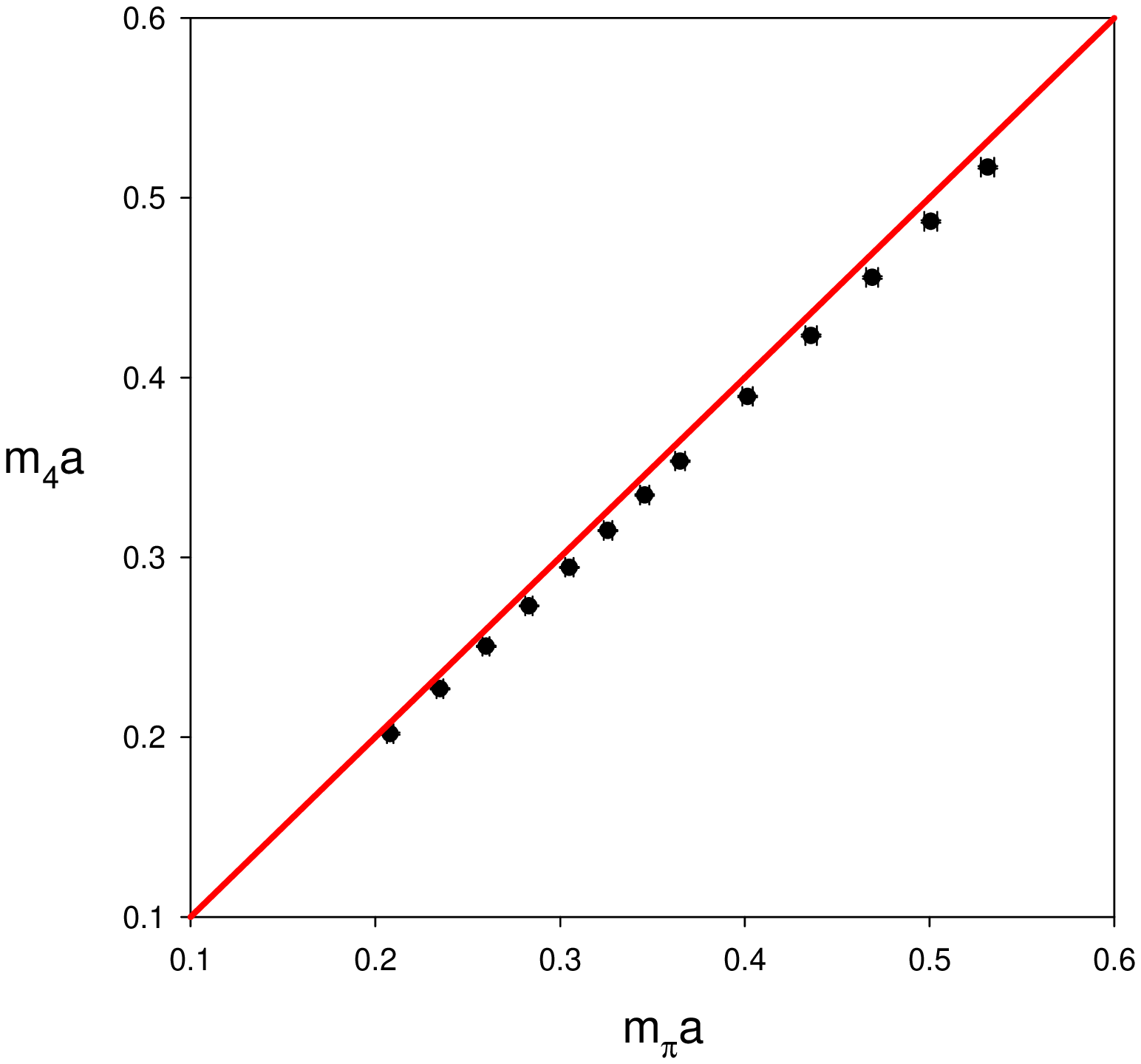}
\caption{The ``pion" mass $ m_4 a $ versus
the pion mass $ m_{\pi} a $. The solid line is $ m_4 = m_{\pi}$.}
\label{fig:m4_mpi}
\end{center}
\end{figure}

The distribution of the indices of $ D $ for these
100 gauge configurations is listed in Table \ref{table:index}.
So the index (topological) susceptibility is
\bea
\label{eq:index_s}
a^4 \chi_t &=&
\frac{ \left< ( n_{+} - n_{-} )^2 \right> }{N} = 6.03(75) \times 10^{-5} \\
\chi_t &=& ( 175 \pm 6 \mbox{ MeV} )^4
\eea
where $ N $ is the total number of sites.

Now substituting the index susceptibility (\ref{eq:index_s}),
the lattice spacing (\ref{eq:a}), $ f_\pi = 132 $ MeV,
and $ N_f = 3 $, into the exact relation (\ref{eq:exact_wv}), we
obtain the $ \eta' $ mass
\bea
m_{\eta'} =  813 \pm 51 \mbox{ MeV} \ ,
\eea
which agrees with the theoretical estimate (\ref{eq:eta_cpt}).
Next we substitute (\ref{eq:index_s}) and
$ f_{\pi} a = 0.0667(2) $ into (\ref{eq:delta_s}), and get
\bea
\label{eq:delta_is}
\delta = 0.16 \pm 0.02 \ ,
\eea
which is in good agreement with the value (\ref{eq:delta_pi})
determined from the data of $m_\pi^2 $, as well as with the
theoretical estimate (\ref{eq:delta_cpt}).

Note that a finite lattice must impose a lower bound for
the pion mass, as well as the corresponding one for the bare quark mass.
If one decreases the bare quark mass beyond its lower bound, then the
resulting pion mass would be flattening (or increasing) rather than
decreasing. This is essentially due to the finite volume effects of the
zero modes of the lattice Dirac operator, which
is proportional to $ |n_{+} - n_{-}|/( m_q^2 a^2 N ) $ as $ m_q a \to 0 $.
Thus, only in the infinite volume limit ($ N \to \infty $),
one can obtain zero pion mass with zero bare quark mass.
Nevertheless, for a finite lattice, the finite volume effects of
the zero modes can be suppressed if the bare quark mass
$ m_q a \gg \sqrt{| n_{+} - n_{-} | / N } $.
For a lattice of size $ 16^3 \times 32 $ at $ \beta = 6.0 $,
we find that $ m_q a \ge 0.03 $ is sufficient to suppress the
finite volume effects of the zero modes.
This can be verified by extracting the ``pion" mass $ m_{4} a $ from the
correlation function of the ``pion" propagator with
$ \gamma_4 \gamma_5 $ coupling
\BAN
\tr[ \gamma_4 \gamma_5 (D_c + m_q)^{-1}(0,x)
     \gamma_4 \gamma_5 (D_c + m_q)^{-1}(x,0) ] \ ,
\EAN
which is void of the contributions of topological zero modes, and
then compare $ m_4 a $ with $ m_{\pi} a $.
In Fig. \ref{fig:m4_mpi}, $ m_4 a $ is plotted against $ m_{\pi} a $,
for the same $ m_q a $.
The good agreement between $ m_4 a $ and $ m_{\pi} a $ for each of the
13 bare quark masses provides an estimate that the error due to
the finite volume effect of the zero modes is less than $ 5\% $,
thus is well under control.

\begin{figure}[htb]
\begin{center}
\hspace{0.0cm}\includegraphics*[height=10cm,width=10cm]{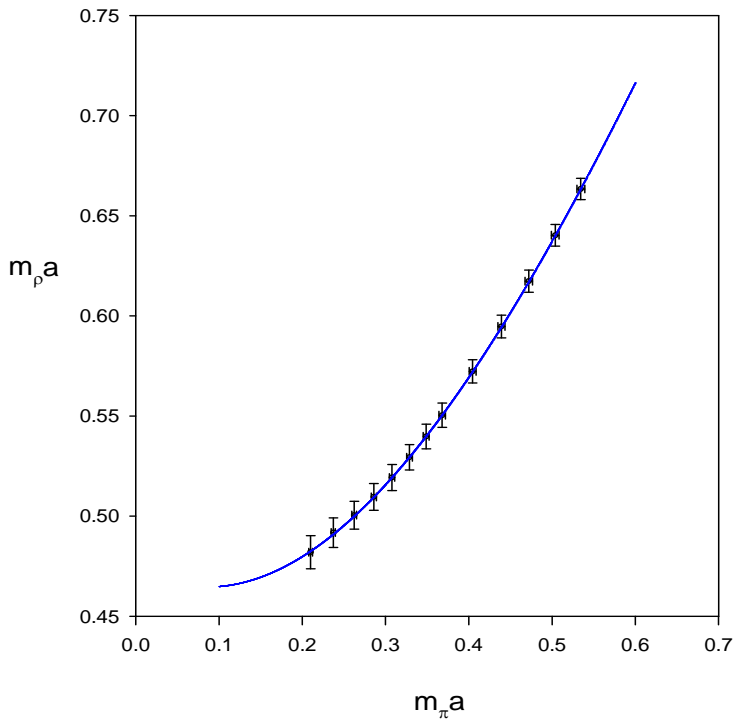}
\caption{The vector meson mass $ m_\rho a $ versus
the pion mass $ m_{\pi} a $ for 13 bare quark masses.
The solid line is the fit to Eq. (\ref{eq:mrho_mpi}). }
\label{fig:mrho_mpi}
\end{center}
\end{figure}

In passing, we present our data of the vector meson mass.
In Fig. \ref{fig:mrho_mpi}, we plot $ m_{\rho} a $ versus $ m_{\pi} a $.
At the smallest bare quark mass $ m_q a = 0.03 $ in our data,
the ratio $ m_\pi / m_{\rho} $ is 0.435(8).
Fitting our data to the vector meson mass formula \cite{Booth:1996hk}
in q$\chi$PT,
\bea
\label{eq:mrho_mpi}
m_\rho a = C_0 + \delta C_{1/2} (m_{\pi} a) + C_1 (m_{\pi} a )^2
           + C_{3/2} (m_{\pi} a)^3 \ ,
\eea
we obtain
\BAN
C_0 &=& 0.474(6)                \\
\delta C_{1/2} &=& -0.223(56)   \\
C_1 &=& 1.370(158)              \\
C_{3/2} &=& -0.542(141)
\EAN
with $ \chi^2 $/d.o.f. = 0.92.

\begin{figure}[htb]
\begin{center}
\hspace{0.0cm}\includegraphics*[height=10cm,width=10cm]{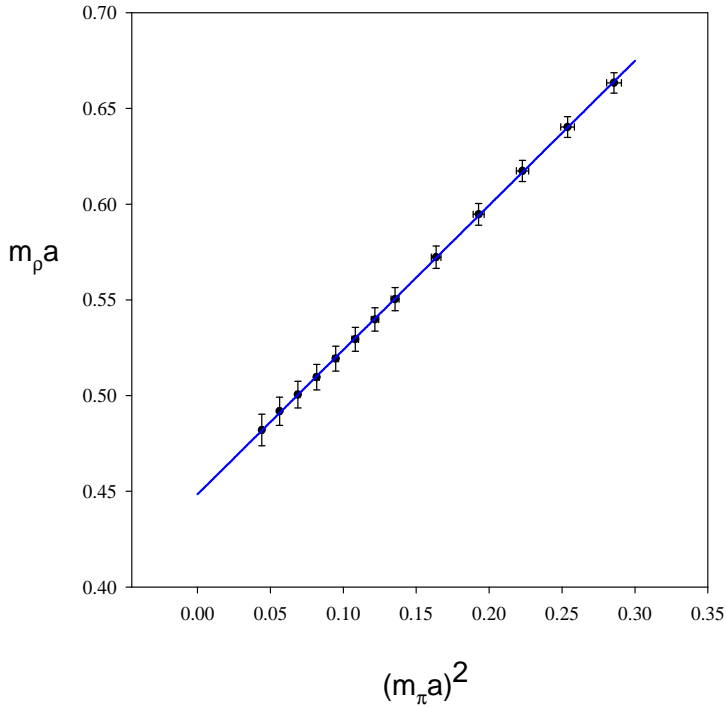}
\caption{The vector meson mass $ m_\rho a $ versus
the pion mass square $ (m_{\pi} a)^2 $ for 13 bare quark masses.
The solid line is the linear fit, Eq. (\ref{eq:mrho_mpi2}). }
\label{fig:mrho_mpi2}
\end{center}
\end{figure}

On the other hand, if we plot $ m_{\rho} a $ versus $ (m_{\pi} a)^2 $,
as shown in Fig. \ref{fig:mrho_mpi2},
then the data seems to be well fitted by
\bea
\label{eq:mrho_mpi2}
m_\rho a = 0.495(4) + 4.57(23) \times (m_{\pi} a)^2
\eea
with $ \chi^2$/d.o.f. = 0.98.

With our present data, it seems to be unlikely
to rule out either one of above two possibilities.
We hope to return to this problem elsewhere.

Note that even though one has no technical difficulties to use smaller 
quark masses to push $ m_\pi/m_\rho $ down to smaller values, 
say, 0.2, one suspects that the finite size effects (especially for $m_\pi$)
might be too large to be meaningful.
Further, for consistency, all quantities in this paper are determined with 
the same set of quark propagators computed for 13 bare quark masses ranging 
from $ m_q a = 0.03 $ to 0.2.

\section{Light Quark Masses}

In the standard model, the quark masses are fundamental parameters which
have to be determined from high energy experiments. However, they cannot
be measured directly since quarks are confined inside hardrons, unlike
an isolated electron whose mass and charge both can be measured directly
from its responses in electric and magnetic fields.
Therefore, the quark masses can only be determined by comparing
a theoretical calculation of physical observables (e.g., hadron masses)
with the experimental values. Evidently, for any field theoretic
calculation, the quark masses depend on the regularization,
as well as the renormalization scheme and scale.

One of the objectives of lattice QCD is to compute the hadron masses
nonperturbatively from the first principle, and from which
the quark masses are determined.
However, the performance of the present generation of computers is still
quite remote from what is required for computing the light hadron masses
at the physical (e.g., $ m_\pi \simeq 140 $ MeV) scale, on a lattice
with enough sites in each direction such that the discretization
errors as well as the finite volume effects are
both negligible comparing to the statistical ones.
Nevertheless, even with lattices of moderate sizes, lattice QCD
can determine the values of the parameters in the hadron mass
formulas of the (quenched) chiral perturbation theory.
Then one can use these formulas to evaluate the hadron masses at the
physical scale, as well as to determine the quark masses.

In quenched chiral perturbation theory \cite{Sharpe:1992ft,Bernard:1992mk},
the ratio of pion and kaon masses to one-loop order reads
\bea
\label{eq:mk2}
\frac{m_K^2}{m_\pi^2} &=& \frac{m+m_s}{2m}
\left\{1+\delta\left[1-\frac{m}{m_s-m}\mbox{ln} \left( \frac{m_s}{m} \right)
\right] \right\}
\eea
where $ m $ denotes $ u $ and $ d $ quark bare masses in the isospin
limit ($ m_u = m_d \equiv m $), $ m_s $ the $ s $ quark bare mass,
$ \delta $ the coefficient of the quenched chiral logarithm.


With experimental values of pion and kaon masses as inputs to
(\ref{eq:mpi2_m}) and (\ref{eq:mk2}), one can determine the ratio of
light quark bare masses $ m_s / m $, but the absolute scale cannot
be fixed by quenched chiral perturbation theory. At this point,
lattice QCD plays the important role to fix the values of the parameters
$ A $, $ B $ and $ \delta $ in the pion mass formula (\ref{eq:mpi2_c}), 
\bea
\label{eq:mpi2_m}
m_\pi^2 = A m^{\frac{1}{1+\delta}} + B m^2 \ ,  
\eea
by measuring the pion mass versus the bare quark mass.
Then the light quark masses $ m $ and $ m_s $
can be determined with experimental inputs of $ m_\pi $, $ m_K $, and
$ f_\pi $ (to fix the lattice spacing $ a $).

In Section 2, we have determined the lattice spacing $ a $, and
the parameters $ \delta $, $ A_1 $ and $ B $ (\ref{eq:a})-(\ref{eq:B_c}),
by computing the quenched quark propagators
for 100 gauge configurations generated with the Wilson gauge action
at $ \beta = 6.0 $ on the $ 16^3 \times 32 $ lattice.
Now inserting $ \delta $ (\ref{eq:delta_pi}) and experimental values
of meson masses ($ m_K = 495 $ MeV and $ m_\pi = 135 $ MeV)
into Eq. (\ref{eq:mk2}), we obtain the quark mass ratio
\bea
\label{eq:ms_m}
\frac{m_s}{m} = 22.58(23) \ ,
\eea
comparing with the ratio at the zeroth order ($ \delta = 0 $)
\bea
\label{eq:ms_m_0}
\left( \frac{m_s}{m} \right)_0 = 25.89
\eea

From (\ref{eq:mpi2_m}), with values of $ a $, $ \delta $,
$ A_1 $ and $ B $ in (\ref{eq:a})-(\ref{eq:B_c}),
and $ m_\pi = 135 $ MeV, we obtain
\bea
\label{eq:m}
m = 4.7 \pm 0.5 \ \mbox{MeV} \ ,
\eea
which is inserted into (\ref{eq:ms_m}) to yield
\bea
\label{eq:ms}
m_s = 107 \pm 11 \ \mbox{MeV} \ .
\eea

In order to transcribe above results (\ref{eq:m})-(\ref{eq:ms})
to their corresponding values in the usual renormalization scheme
$ \overline{\mbox{MS}} $ in high energy
phenomenology, one needs to compute the lattice renormalization
constant $ Z_m = Z_s^{-1} $, where $ Z_s $ is the renormalization
constant for $ \bar{\psi} \psi $. In general, $ Z_m $ should be
determined non-perturbatively. In this paper, we use the one loop
perturbation formula \cite{Alexandrou:1999wr}
\bea
\label{eq:Zs}
Z_s = 1 + \frac{ g^2 }{ 4 \pi^2 }
\left[ \mbox{ln} ( a^2 \mu^2 ) + 6.7722 \right], \hspace{8mm}
(m_0=1.30) \ ,
\eea
to obtain an estimate of $ Z_s $.
At $ \beta = 6.0 $, $ a = 0.0997(3) $ fm, and $ \mu = 2 $ GeV, (\ref{eq:Zs})
gives $ Z_s = 1.172(1) $. Thus the quark masses in (\ref{eq:m})-(\ref{eq:ms})
are transcribed to
\bea
\label{eq:m_MS}
m_{u,d}^{\overline{\mbox{MS}}}( 2 \mbox{ GeV} )
&=& 4.1 \pm 0.3 \pm 1.0 \mbox{ MeV }  \\
\label{eq:ms_MS}
m_s^{\overline{\mbox{MS}}}( 2 \mbox{ GeV} )
&=& 92 \pm 9 \pm 16 \mbox{ MeV}
\eea
where the first errors are statistical errors, and the second errors
are our estimate of systematic errors due to finite volume and
discretization effects, by comparing our present results to
our earlier ones \cite{Chiu:2002rk} with smaller volume and
larger lattice spacing. Evidently, the light quark masses in
(\ref{eq:m_MS})-(\ref{eq:ms_MS}) are in good agreement with
the current lattice world average \cite{Wittig:2002ux}, as well 
as the strange quark mass determined in Ref. \cite{Giusti:2001pk}.


\section{Chiral Condensate and Quark-Gluon Condensate}

\subsection{Chiral Condensate}

The chiral condensate in lattice QCD with the optimal DWF
can be expressed as
\bea
\label{eq:qbq}
-\langle \bar q q \rangle &=&
 \lim_{m_q \to 0} \lim_{\Omega \to \infty}
 \frac{1}{\Omega} \sum_x \left.\frac{\delta}{\delta J(x)}
 \frac{\delta}{\delta \bar J(x)} Z[J, \bar J] \right|_{J=\bar J=0} \nn
&=& \lim_{m_q \to 0} \lim_{\Omega \to \infty} \frac{1}{\Omega}
\frac{\int [dU] \ e^{-{\cal A}_g} \ \det D(m_q) \
      \sum_x \tr(D_c+m_q)^{-1}_{x,x}}
     {\int [dU] \ e^{-{\cal A}_g} \ \det D(m_q) }  \nn
&\equiv& \Sigma(N_f)
\eea
where $ \Omega = Na^4 $ is the volume of the 4D lattice,
$ N_f $ is the number of quark flavors, 
$ Z[J, \bar J] $ is the generating functional in (\ref{eq:ZW_odwf}),
and the Dirac, color, and flavor indices have been suppressed.

The subtlety of (\ref{eq:qbq}) lies in its two limiting processes.
If the order of these two limiting processes is exchanged, then
$ \langle \bar q q \rangle $ must be zero since spontaneous chiral
symmetry could not occur on a finite lattice for exactly massless
fermions. 
Thus it seems to be unlikely to be able to extract the
chiral condensate of QCD on a finite lattice, not to mention
in the quenched approximation\footnote{
Recall that the $ \eta' $ loop in the quark propagator leads to the
quenched chiral logarithm in the chiral condensate.}.
Nevertheless, using chiral perturbation theory or random matrix theory, 
the chiral condensate at finite volume and quark mass can be computed 
in the continuum, both for unquenched and quenched QCD. 
In the quenched approximation, 
the chiral condensate with fixed $ m_q $, $ \Omega $, 
and $ Q $ (topological charge) reads \cite{Osborn:1998qb}
\bea
\label{eq:qbq_q} 
-\langle \bar q q \rangle_{m,\Omega,Q}
= \Sigma \left\{ \frac{|Q|}{z} + 
  z \left[ I_{|Q|}(z) K_{|Q|}(z) + I_{|Q|+1}(z) K_{|Q|-1}(z) \right] \right\} 
\eea
where $ \Sigma = \Sigma(N_f=0) $, $ z = m_q \Omega \Sigma $, and 
$ I_Q $ and $ K_Q $ are modified Bessel functions. 
Thus, the chiral condensate can be extracted by measuring 
$ \langle \bar q q \rangle $ in different topological 
sectors at different $ m_q $ and $ \Omega $. 
Alternatively, a plausible way to extract the chiral condensate 
on a finite lattice (with large volume $ \Omega $) is to measure 
$ \tr(D_c+m_q)^{-1}_{x,x} $, i.e.,   
\bea
\label{eq:qbq_1}
-\langle \bar q q \rangle =
\frac{\int [dU] \ e^{-{\cal A}_g} \ \tr(D_c+m_q)^{-1}_{x,x}}
     {\int [dU] \ e^{-{\cal A}_g} } \ ,  
\eea
for a set of small quark masses $ m_q $ satisfying $ z \gg 1 $.   
Then the chiral condensate at $ m_q = 0 $ can be obtained by linear 
extrapolation, since $ \tr(D_c+m_q)^{-1}_{x,x} $
behaves like a linear function of $ m_q $ for small $ m_q $. 
This ansatz can be verified in practice.

\begin{figure}[htb]
\begin{center}
\hspace{0.0cm}\includegraphics*[height=10cm,width=10cm]{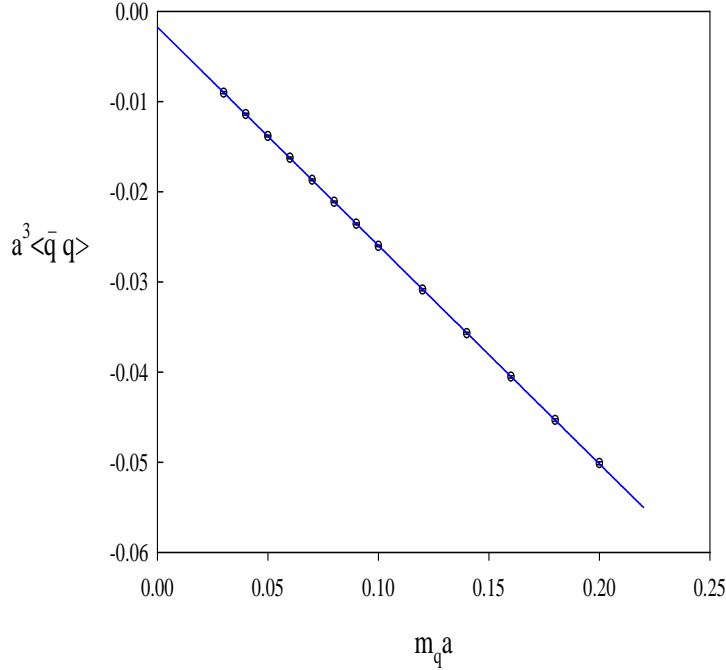}
\caption{The chiral condensate $ a^3 \langle \bar q q \rangle $ versus
the bare quark mass $ m_q a $. The solid line is the linear
fit (\ref{eq:qbq_fit}).}
\label{fig:qbq}
\end{center}
\end{figure}

In the following, we measure $ \langle \bar q q \rangle $ in the
quenched approximation, with one fixed site at the origin
$ (\vec{0},0) $, for 100 gauge configurations generated with
Wilson gauge action at $ \beta = 6.0 $ on the $ 16^3 \times 32 $ lattice.
The results are plotted in Fig. \ref{fig:qbq} for 13 bare quark masses,
and they are very well fitted by the straight line
\bea
\label{eq:qbq_fit}
-a^3 \langle \bar q q \rangle =
1.73(3) \times 10^{-3} + 0.242(0) \times (m_q a) \ .
\eea
Now we take the value $ 1.73(3) \times 10^{-3} $ at $ m_q a = 0 $ to be
the chiral condensate, then we obtain
\bea
\label{eq:qbq_lat}
\langle \bar q q \rangle = -0.0134(2) \mbox{ GeV}^3
\eea
with the inverse lattice spacing $ a^{-1} $ (\ref{eq:a}) determined
in Section 2. Using the one loop renormalization $ Z_s $ (\ref{eq:Zs}),
we transcribe the chiral condensate (\ref{eq:qbq_lat}) to
$ \overline{\mbox{MS}} $ at scale $ \mu = 2 \mbox{ GeV} $,
\bea
\label{eq:qbq_MS}
\langle \bar q q \rangle^{\overline{\mbox{MS}}}(2 \mbox{ GeV})
= -(250 \pm 3 \mbox{ MeV})^3 \ ,
\eea
which is in good agreement with the phenomenological estimate,
as well as other recent lattice results
\cite{Giusti:2001pk,Hasenfratz:2002rp,Hernandez:2001hq}.
Nevertheless, a complete picture of the underlying mechanism
of spontaneous symmetry breaking in QCD can only be unveiled
in the context of dynamical quarks.

\subsection{Quark-Gluon Condensate}

The quark-gluon condensate
$ g \langle \bar q \sigma_{\mu\nu} F_{\mu\nu} q \rangle $
is one of the nonperturbative features of the QCD vacuum.
It is one of the parameters in the framework
of QCD sum rule to determine the low-energy properties of QCD
\cite{Shifman:bx,Ioffe:kw}.

The first measurement \cite{Kremer:1987ve} of the quark-gluon
condensate in lattice QCD was performed more than 15 years ago,
with the staggered fermion, for only 5 gauge configurations.
A recent measurement \cite{Doi:2002wk} was also
carried out with the staggered fermion, at each site along the body
diagonal of the $ 16^4 $ lattice, for 100 gauge configurations generated
with Wilson gauge action at $ \beta = 6.0 $. In Ref. \cite{Doi:2002wk},
with 1600 measurements for each quark mass (of total 3 quark masses),
the quark condensate and quark-qluon condensate at $ m_q = 0 $
were obtained by linear extrapolation, however, the ratio
$ g \langle \bar q \sigma_{\mu\nu} F_{\mu\nu} q \rangle /
  \langle \bar q q \rangle $ turns out to be much larger
  than the value in the QCD sum rule as well as other
  phenomenological estimates.

In general, the quark-gluon condensate in lattice QCD with the
optimal DWF can be expressed as
\bea
\label{eq:pbpg}
& & - g \langle \bar q \sigma_{\mu\nu} F_{\mu\nu} q \rangle \nn
&=& \hspace{-4mm}
\lim_{m_q \to 0} \lim_{\Omega \to \infty}
\frac{1}{\Omega} \sum_x \left.\frac{\delta}{\delta J(x)}
        \sigma_{\mu\nu} F_{\mu\nu}(x)
       \frac{\delta}{\delta \bar J(x)} Z[J, \bar J] \right|_{J=\bar J=0} \nn
&=& \hspace{-4mm}
\lim_{m_q \to 0} \lim_{\Omega \to \infty} \frac{1}{\Omega}
\frac{\int [dU] \ e^{-{\cal A}_g} \ \det D(m_q) \
      \sum_x \tr[(D_c+m_q)^{-1}_{x,x} \sigma_{\mu\nu} F_{\mu\nu}(x)] }
     {\int [dU] \ e^{-{\cal A}_g} \ \det D(m_q) }
\eea
where $ \Omega = (N a)^4 $ is the volume of the 4D lattice,
$ Z[J, \bar J] $ is the generating functional in (\ref{eq:ZW_odwf}),
and the Dirac, color, and flavor indices have been suppressed.
Here the matrix valued field tensor $ F_{\mu\nu}(x) $ can be
obtained from the four plaquettes surrounding $ x $ on the
$ (\hat\mu, \hat\nu) $ plane, i.e.,
\BAN
g a^2 F_{\mu\nu}(x)
\hspace{-2mm}
& \simeq & \hspace{-2mm}
\frac{1}{8i}
   [  P_{\mu\nu}(x) + P_{\mu\nu}(x-\hat\mu) + P_{\mu\nu}(x-\hat\nu)
      + P_{\mu\nu}(x-\hat\mu-\hat\nu)  \nn
& &
      - P^{\dagger}_{\mu\nu}(x) - P^{\dagger}_{\mu\nu}(x-\hat\mu)
      - P^{\dagger}_{\mu\nu}(x-\hat\nu)
      - P^{\dagger}_{\mu\nu}(x-\hat\mu-\hat\nu) ]
\EAN
where
\BAN
P_{\mu\nu}(x) = U_\mu(x) U_\nu(x+\hat\mu)
                U^{\dagger}_\mu(x+\hat\nu) U^{\dagger}_\nu(x) \ .
\EAN

\begin{figure}[htb]
\begin{center}
\hspace{0.0cm}\includegraphics*[height=10cm,width=10cm]{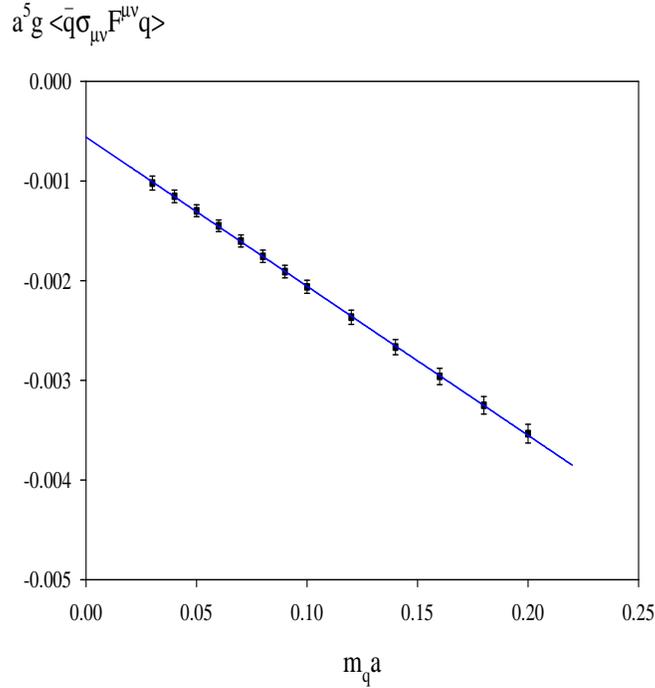}
\caption{The quark-gluon condensate
$ a^5 g \langle \bar q \sigma_{\mu\nu} F_{\mu\nu} q \rangle $ versus
the bare quark mass $ m_q a $. The solid line is the linear fit.}
\label{fig:pbpg1}
\end{center}
\end{figure}

In the following, we measure the quark-gluon condensate in
quenched approximation, with one fixed site at the origin
$ (\vec{0},0) $, for 100 gauge configurations generated with
Wilson gauge action at $ \beta = 6.0 $ on the $ 16^3 \times 32 $ lattice.
The results are plotted in Fig. \ref{fig:pbpg1} for 13
bare quark masses, and they are well fitted by the straight line
\bea
- a^5 g \langle \bar q \sigma_{\mu\nu} F_{\mu\nu} q \rangle
= 5.59(12) \times 10^{-4} + 0.0150(0) \times (m_q a) \ .
\eea
Now we take the value $ 5.59(12) \times 10^{-4} $ at $ m_q a = 0 $ to be
the quark-gluon condensate, then we obtain
\bea
\label{eq:pbpg_lat}
g \langle \bar q \sigma_{\mu\nu} F_{\mu\nu} q \rangle
= -0.0170(4) \mbox{ GeV}^5
\eea
with the inverse lattice spacing $ a^{-1} $ (\ref{eq:a}) determined
in Section 2. Using the one loop
renormalization \cite{Alexandrou:1999wr},
we transcribe the quark-gluon condensate (\ref{eq:pbpg_lat}) to
$ \overline{\mbox{MS}} $ at scale $ \mu = 2 \mbox{ GeV} $,
\bea
\label{eq:pbpg_MS}
g \langle \bar q \sigma_{\mu\nu} F_{\mu\nu} q \rangle^{\overline{\mbox{MS}}}
(2 \mbox{ GeV})
= -(434 \pm 4 \mbox{ MeV})^5 \ ,
\eea
Thus the ratio of the quark-gluon condensate (\ref{eq:pbpg_MS}) to
the quark condensate (\ref{eq:qbq_MS}) is
\bea
\label{eq:M02}
M_0^2 = \left. \frac{g \langle \bar q \sigma_{\mu\nu} F_{\mu\nu} q \rangle}
             {\langle \bar q q \rangle} \right|^{\overline{\mbox{MS}}}
(2 \mbox{ GeV}) = 0.98(2) \mbox{ GeV}^2 \ ,
\eea
which becomes $ 0.92(2) \mbox{ GeV}^2 $ at $ \mu = 0.5 \mbox{ GeV} $,
in agreement with the estimate $ 0.8 \pm 0.2 \mbox{ GeV}^2 $
\cite{Belyaev:sa} in the QCD sum rule.

\section{Summary and Discussions}

In this paper, we have determined several quantities
[the topological susceptibiltiy (\ref{eq:index_s}),
the parameters $ \delta, A_1, B $ (\ref{eq:a})-(\ref{eq:B})
in the pseudoscalar meson mass formulas,
the light quark masses (\ref{eq:m_MS})-(\ref{eq:ms_MS}),
the quark condensate (\ref{eq:qbq_MS}), and
the quark-gluon condensate (\ref{eq:pbpg_MS})]
in quenched lattice QCD with exact chiral symmetry, with quark fields
defined by the boundary modes (\ref{eq:q})-(\ref{eq:qbar})
of the optimal DWF. The propagator
of the 4D effective Dirac operator, $ D^{-1}(m_q) $, (\ref{eq:Dm_RZ})
is solved with Neuberger's 2-pass algorithm \cite{Neuberger:1998jk},
for 100 gauge configurations \cite{Kilcup:1996hp} generated by Wilson
gauge action at $ \beta=6.0 $ on the $ 16^3 \times 32 $ lattice.
Our results provide strong evidence that lattice QCD with exact
chiral symmetry indeed realizes the quenched QCD chiral dynamics.

The coefficient of quenched chiral logarithm
($ \delta = 0.164 \pm 0.013 $)
extracted from the pion mass agrees very well with that
($ \delta = 0.16 \pm 0.02 $) obtained from the topological
susceptibility. The agreement provides a consistency check of
the theory. Also, both of them are in good agreement with the
theoretical estimate $ \delta \simeq 0.176 $ in quenched chiral
perturbation theory.

Even though the size of our lattice is insufficient
for computing the light hadron masses at the physical
(e.g., $ m_\pi \simeq 140 $ MeV) scale,
we can determine the values of the parameters $ \delta, A_1, B $
(\ref{eq:a})-(\ref{eq:B})
in the pseudoscalar meson mass formulas of q$\chi$PT, and then use
these formulas to evaluate the pseudoscalar masses at the physical scale,
as well as to determine the light quark masses $m_{u,d}$ and $m_s$,
(\ref{eq:m_MS})-(\ref{eq:ms_MS}).

Using the Gell-Mann-Oakes-Renner relation,
\BAN
m_\pi^2 = -\frac{ 4 m \left< \bar\psi \psi \right> }{f_\pi^2} \ ,
\EAN
and Eqs. (\ref{eq:mk2}) and (\ref{eq:mpi2}),
we can estimate the chiral condensate at the zeroth order
\bea
\label{eq:ubu_0}
\left< \bar{u} u \right>_0 &=& -\frac{C f_\pi^2}{4}
= -0.00942(18) \mbox{ GeV}^3                         \\
\label{eq:sbs_0}
\left< \bar{s} s \right>_0 &=&
\left< \bar{u} u \right>_0
\frac{m}{m_s} \left( \frac{ 2 m_K^2 }{ m_\pi^2 } - 1 \right)
= -0.0108(3) \mbox{ GeV}^3
\eea
where (\ref{eq:Ca}), (\ref{eq:a}), (\ref{eq:ms_m}), and the experimental
inputs ($ m_\pi = 135 $ MeV, $ m_K = 495 $ MeV) have been used.
Using one-loop renormalization (\ref{eq:Zs}),
we transcribe (\ref{eq:ubu_0}) to
$ \overline{\mbox{MS}} $ at scale $ \mu = 2 \mbox{ GeV} $,
\bea
\label{eq:ubu_0_MS}
\left< \bar{u} u \right>_0^{\overline{\mbox{MS}}}(2 \mbox{ GeV})
&=& -(223 \pm 3 \mbox{ MeV})^3 \ , \\
\label{eq:sbs_0_MS}
\left< \bar{s} s \right>_0^{\overline{\mbox{MS}}}(2 \mbox{ GeV})
&=& -(233 \pm 6 \mbox{ MeV})^3 \ .
\eea
It is expected that the values of the quark condensates (\ref{eq:ubu_0_MS})
-(\ref{eq:sbs_0_MS}) are different from (\ref{eq:qbq_MS}), since
the formers are zeroth order estimates using the pseudoscalar mass formulas
with parameters determined by the long distance behaviors of the
pion propagator, while the latter is determined by the short
distance bahaviors of the quark propagator.

The quark-gluon condensate as well as the chiral condensate
are the basic features of the QCD vacuum. In the framework
of the QCD sum rules, they are introduced as parameters in the
nonperturbative generalization of operator product expansion,
and are related to hardonic properties. Therefore, it is
important for lattice QCD to determine these quantities from
the first principles. Our determination of the quark-gluon condensate
(\ref{eq:pbpg_MS}) is the first result of this quantity from
lattice QCD with exact chiral symmetry.
The agreement of the ratio
(of the quark-gluon condensate to the chiral condensate),
$ {M_0^2}_{\overline{\mbox{MS}}}(0.5 \mbox{ GeV}) = 0.92(2) \mbox{ GeV}^2 $,
with the estimate $ 0.8(2) \mbox{ GeV}^2 $ in the QCD sum rule is
encouraging, which suggests that lattice QCD with exact chiral symmetry
does provide a viable framework to tackle the low energy physics of QCD
pertaining to the nonperturbative QCD vacuum.


\bigskip
\bigskip
\flushpar
{\bf Acknowledgement }
\bigskip

\noindent

T.W.C. would like to thank Steve Sharpe for helpful discussions 
and suggesting (\ref{eq:mpi2_c}).
This work was supported in part by the National Science Council,
ROC, under the grant number NSC91-2112-M002-025.


\vfill\eject

\end{document}